# Multimodal Assessment of Speech Impairment in ALS Using Audio-Visual and Machine Learning Approaches

*Francesco Pierotti*[1], *Andrea Bandini*[1,2,3]

[1]The BioRobotics Institute and Department of Excellence in Robotics and AI, Scuola Superiore Sant'Anna, Pisa, Italy
[2]Health Science Interdisciplinary Research Center, Scuola Superiore Sant'Anna, Pisa, Italy
[3]KITE Research Institute, University Health Network, Toronto, ON, Canada

```
Francesco.pierotti@santannapisa.it
```

## Abstract

The analysis of speech in individuals with amyotrophic lateral sclerosis is a powerful tool to support clinicians in the assessment of bulbar dysfunction. However, current methods used in clinical practice consist of subjective evaluations or expensive instrumentation. This study investigates different approaches combining audio-visual analysis and machine learning to predict the speech impairment evaluation performed by clinicians. Using a small dataset of acoustic and kinematic features extracted from audio and video recordings of speech tasks, we trained and tested some regression models. The best performance was achieved using the extreme boosting machine regressor with multimodal features, which resulted in a root mean squared error of 0.93 on a scale ranging from 5 to 25. Results suggest that integrating audio-video analysis enhances speech impairment assessment, providing an objective tool for early detection and monitoring of bulbar dysfunction, also in home settings.

## 1. Introduction

Amyotrophic lateral sclerosis (ALS) is a rare neurodegenerative disease affecting 4.42 per 100,000 people globally, with an estimated survival time of 3 to 5 years after the disease onset [1], [2]. However, survival time can vary depending on several factors including the age and site of onset (e.g., bulbar or spinal) and the rate of disease progression. ALS is characterized by the degeneration of motor neurons in the brain and spinal cord, causing muscle weakness and paralysis in the limbs, head, and neck. This degeneration often causes speech and swallowing disorders [3]. The loss of speech function is a hallmark of reduced quality of life for individuals with ALS [4].

Due to the heterogeneous nature of disease progression, identifying biomarkers of prognosis and developing predictive models remain crucial challenges. Early and accurate assessment of bulbar impairment and its decline is essential for gaining valuable insight into disease progression and for rapidly accessing therapeutic strategies, improving the effectiveness of treatment outcomes [5].

The assessment of bulbar impairment in ALS is particularly challenging because it often relies on subjectively reported symptoms [6], which can result in delays in both diagnosis and prognosis [7]. Currently, the assessment of compromised orofacial functions caused by bulbar impairment largely depends on evaluations conducted by speech-language pathologists (SLPs) or on objective sensor-based measurements [8]. However, these methods require expensive instrumentation such as videofluoroscopy and fiberoptic endoscopy, and effort, which may limit the accessibility to an appropriate diagnosis. Moreover, traditional methods rely on in-person visits, which can be costly and logistically complicated for those living far from socialized medical centres. Additionally, due to its rarity, ALS represents a challenge for collecting sufficiently large datasets. This highlights the urgent need for objective, cost-effective, and non-invasive diagnostic methods to overcome these limitations.

Multicentric studies are crucial for acquiring a more representative sample of the population. However, a multicentric study may introduce logistical and coordination challenges, which may limit the effectiveness of their goal. To overcome these limitations in data collection, remote monitoring has emerged as a promising solution. Indeed, audio and video recording enable the data collection directly from their own homes, reducing the burden of travelling, and allowing a more frequent and naturalistic assessment of speech impairment [9].

In recent years, speech acoustics and orofacial kinematics have shown significant potential in addressing these challenges, paving the way for simpler and more accessible methods to assess bulbar function [10]. These approaches could significantly reduce the costs and effort required for both individuals and the healthcare system [11], [12]. Specifically, acoustic analysis of impaired speech functions has emerged as a promising approach for the assessment and monitoring of ALS, with several speech metrics capable of describing functional aspects of bulbar impairment [13], [14], [15]. Indeed, Simmatis et al, [15] employed acoustic analysis to detect ALS and assess speech impairment severity. Additionally, several studies utilizing video-based analysis to extract orofacial kinematics [3], [16], [15] have demonstrated the feasibility of remotely and continuously evaluating speech functions. For example, Bandini et al. [3] proposed a markerless video-based approach to discriminate ALS individuals from neurotypical controls. Furthermore, some researchers studied multimodal approaches [11], [12], [18], [19], combining acoustic and visual data to enhance the assessment accuracy and reliability, In this context, Neumann et al. [11] investigated the utility of multimodal measures for the assessment and monitoring of ALS. Moreover, advancements in artificial intelligence (AI) and computer vision have further enhanced the feasibility and accuracy of these methods, with several studies highlighting the potential of automated tools for ALS assessment [20], [21].

While significant advancements have been made in the assessment of ALS speech impairment using acoustic and kinematics features of speech, these approaches mainly rely on classification models to distinguish between ALS and control individuals, or on the description of suitable biomarkers for the disease progression. Therefore, these methods do not provide a fine-grained estimation of clinical outcomes, which is necessary to keep track of disease progression and evaluate treatment efficacy.

In this study, we addressed this gap by adopting a multimodal approach that combines clinical evaluations, AI, and audio-visual analysis to predict bulbar function in ALS. Our approach aims to quantitively estimate the speech impairment severity assessed by clinicians, and to evaluate the efficacy of such a multimodal approach compared to single-modal approaches, providing a comprehensive and accurate framework for assessing bulbar impairment in ALS.

## 2. Methods

### 2.1. Dataset description

In this study, we used the Toronto NeuroFace dataset [21], which includes 11 individuals with ALS (6 females, age range 45-75 years) and 11 individuals with no speech and orofacial impairments as a control group (HC) (4 females, age range 33-78 years).

The dataset contains audio and video recordings of participants performing various speech and non-speech tasks typically conducted during speech and orofacial function assessments (sentence repetition, diadochokinesis of syllables /pa/ and /pataka/, lip puckering, smiling, maximum opening of the jaw, and raising of the eyebrows). Additionally, researchers provided demographic and clinical information about the recruited groups, along with an assessment of orofacial impairment by two trained SPLs with a score on a scale from 1 (normal function) to 5 (severe dysfunction). These scores were evaluated based on symmetry, range of motion, speed, variability, and movement-related fatigue. The sum of the five scores was used as the total score.

### 2.2. Data analysis

Figure 1 shows the pipeline adopted in this study. Among all the tasks included in the dataset, we conducted analyses only on the audio and video of the sentence "Buy Bobby a puppy" (BBP), which was repeated 10 times at a comfortable speaking rate and loudness, with a brief pause between each repetition.

Each BBP repetition was manually segmented on both streams by visually inspecting the audio and video recordings. Specifically, we identified the onset and offset timings of each sentence by analyzing the minimum energy points in the audio signals. For each frame of the video recordings, we first identified the facial region of the participant using the single shot scale-invariant face detector (SFD) face detection model [23], and then we detected the facial landmarks using the face alignment network (FAN) algorithm [24], able to accurately estimate the 3D coordinates of 68 facial landmarks based on four stacked hourglass CNNs. These methods have been shown to perform accurately in face alignment across different groups of individuals with neurological disorders [22]. The 68 landmark points were extracted from each frame, estimating the position of eyes, eyebrows, nose, mouth, and jawline. To segment the video into individual repetitions, we manually inspected video frame by frame, annotating the frames relative to the onset and offset of each sentence.

### 2.3. Feature extraction

Several audio features, including phonatory, prosodic, and intelligibility features, were extracted from each repetition. Audio features were extracted using Parselmouth [25], a Python library for the PRAAT software [26], and the SpeechRecognition library, by exploiting also the functions vosk API. Specifically, the extracted features included metrics of the fundamental frequency (F0), jitter, shimmer, harmonics-to-noise ratio, total sentence duration, inter-sentence duration, pause duration, word error rate, and dynamic time-warping (DTW). Since the DTW feature for the audio recordings was computed using the first repetition as a template, that repetition was excluded from all participants in the subsequent analyses. These acoustic features provide insights into the speech production system [27]. For example, the phonatory features (F0, jitter, shimmer, and HNR) are indicators of phonatory stability, which may be impacted in ALS due to vocal fold dysfunction [28]. Similarly, prosodic features reflect speech timing and fluency [29].

Additionally, video features capturing range of motion, speed, symmetry, and shape were extracted from the facial landmarks of each frame. Specifically, these features included the cumulative path travelled by the lower lip and by the jaw, the mean value and the range of the mouth area with respect to the rest position, the maximum and minimum velocities of the mouth width, lower lip, and jaw, the jaw jerk, the absolute difference between right and left mouth areas, the correlation between right and left mouth corners movements, the mean value of mouth eccentricity, and the range of mouth eccentricity. These features have been demonstrated to be highly sensitive during speech due to impaired muscle strength, speed, and coordination [16]. To ensure consistency across frames, all calculations were normalized by using the intercanthal distance. Overall, 33 features were extracted, with 15 coming from videos and 18 from audio.

For each modality (audio-based, video-based, and multimodal), the extracted features were analyzed using the same pipeline. The target value to be estimated with our proposed approach was the mean value of the total scores assigned by the two SLPs, which ranges between 5 and 25

### 2.4. Regression

In this phase, we trained and tested a support vector regression (SVR) model, a multi-layer perceptron (MLP) regressor neural network, and an extreme boosting machine (XGB) regressor to automatically estimate the level of speech impairment. Due to the relatively low number of participants, these models were trained and evaluated using a nested leave-one-subject-out cross-validation to identify the most suitable model for each participant. For each iteration, all repetitions of one participant were used as the test set, while instances from the other participants served as the training and validation set. At each iteration, all features were standardized using the mean values and standard deviations of the current train set. The scores were predicted using the best configuration of the hyperparameters, optimized on the training instances using another leave-one-subject-out cross-validation. To determine the best configuration, a grid search of the hyperparameters was performed for each model.

Therefore, the performance of the regression approaches was compared in terms of root mean square error (RMSE) with respect to the speech impairment scores provided by the trained speech-language pathologists. Using a subject-based approach, an RMSE was computed for each participant considering their own repetitions. Then, the mean RMSE (mRMSE) was computed as the average of all the participant RMSEs. Finally, the coefficients of variation (CV) of the RMSE of ALS and HC groups were calculated to study the stability of the models based on the group under analysis.

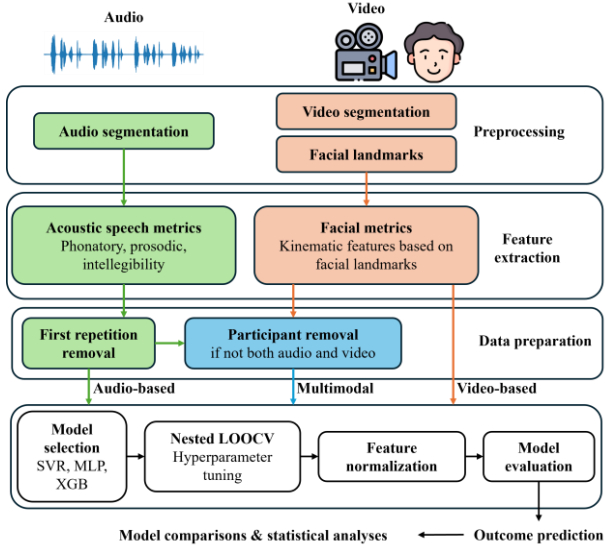

Figure 1: *schematic pipeline of the audio-video assessment of speech impairment via regression of the level of speech impairment.*

### 2.4.1. Hyperparameter optimization

For SVR the grid search was performed to optimize the combination of hyperparameters $C$ and $\varepsilon$ (with $C = [10^{-1}, 1, 10, 10^2, 10^3, 10^4]$, $\varepsilon = [0.01, 0.1, 0.5, 1]$). The kernel type was also tuned, selecting the best one among linear, radial basis function, and sigmoid.

For MLP, hyperparameters were tuned for the number of hidden layers and nodes per layer ([(10, 50), (10, 30, 100), (10, 50, 100), (10, 50, 200), (10, 100, 100), (10, 100, 200), (50, 10), (100, 30, 10), (100, 50, 10), (200, 50, 10), (100, 100, 10), (200, 100, 10)]), learning rate $\alpha = [0.0001, 0.001, 0.01]$, and activation function ('*identity*', '*logistic*', '*tanh*', '*relu*').

For the XGB, the grid search optimized combination of number of estimators (2, 3, 4, 5), maximum depth (3, 4, 5, 6, 8), learning rate (0.001, 0.05, 0.01, 0.1, 0.15, 0.3), subsample (0.1, 0.3, 0.5, 0.7, 1.0), and column sample by tree (0.1, 0.3, 0.5, 0.7, 1.0).

### 2.4.2. Statistical anlysis

To study significant differences among the results, we first verified the normality distribuition of the RMSEs with the Shapiro-Wilk test, and the we applied the Friedman's test to study any significant effects of the modality (audio, video, and audio-video) and regression models.

## 3. Results

For the multimodal approach, we considered only participants who had both audio and video recordings. Additionally, due to some inconsistencies in the number of repetitions between the two modalities for some participants, we carefully removed instances that were present in only one modality. As a result, the number of participants included in the video-based analysis was 11 HC and 9 ALS (for a total of 202 instances) while in the audio-based and multimodal analysis were 9 HC and 8 ALS (for a total of 153 and 151 instances, respectively).

The best performance in predicting the speech impairment scores was achieved by SVR regressor trained on the combination of audio and video features, yielding a mRMSE of 0.93 (1.29 for ALS and 0.62 for HC), as shown in Figure 2. For the audio features analysis, the best results were obtained using XGB regressor, achieving an overall mRMSE of 0.99 (1.23 for ALS and 0.77 for HC). In contrast, the video-based approach achieved an mRMSE of 1.21 (1.75 for ALS and 0.77 for HC) with XGB regressor. The mRMSE of each regressor in each experiment is reported in Table 1.

As can be seen form Table 1, for all models in each modality, the mRMSE of the HC group is lower than the mRMSE of the ALS group.

Table 1: *Performance of regressors in different modes (CV in brackets).*

| Modality | Model | mRMSE | mRMSE (ALS) | mRMSE (HC) |
|---|---|---|---|---|
| Audio | SVR | 1.08 | 1.41 (0.65) | 0.80 (0.47) |
| | MLP | 1.09 | 1.46 (0.57) | 0.76 (0.51) |
| | XGB | 0.99 | 1.23 (0.81) | 0.77 (0.46) |
| Video | SVR | 1.21 | 1.75 (0.77) | 0.77 (0.64) |
| | MLP | 1.37 | 1.88 (0.69) | 0.96 (0.51) |
| | XGB | 1.30 | 1.73 (0.71) | 0.95 (0.44) |
| Audio-Video | SVR | 0.93 | 1.29 (0.94) | 0.62 (0.28) |
| | MLP | 1.05 | 1.33 (0.68) | 0.80 (0.49) |
| | XGB | 1.06 | 1.28 (0.72) | 0.86 (0.29) |

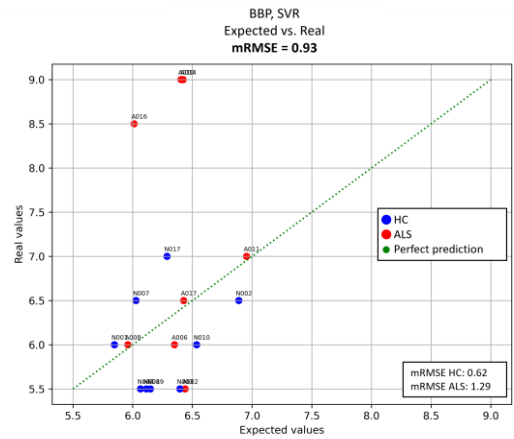

Figure 2: *SVR regressor with multimodal features. ALS participant scores are highlighted in red, HC scores in blue.*

As shown in Figure 2, the ALS participants are almost perfectly aligned with the perfect prediction line, and those having the greater distances from the line, are the ones with higher real scores. These participants have predicted scores lower than the actual ones. Similarly, HC participants are almost perfectly aligned with a prediction error (0.62) lower than that of ALS participants (1.29).

No statistically significant effects (p=0.092, $\chi^2$ =13.6267, df=8) were observed for the models or modality in the statistical analysis.

## 4. Discussion

In this study, we compared different approaches for predicting speech impairment scores in individuals with ALS. We aimed to investigate whether a multimodal approach based on audio and video features is feasible and can further improve clinical outcome predictions.

The overall mRMSE of 0.93 achieved by the SVR regressor is a promising result for the multimodal approach. Moreover, for individuals with low or moderately low speech impairment scores, the model accurately predicts the clinical outcome; however, for higher scores, it tends to be more conservative. Specifically, ALS participants with more severe impairment tended to have predictions above the perfect prediction line in Figure 2, with greater deviations from it compared to those with lower impairment levels. This poorer prediction for participants with a clinical outcome above or equal to 8.5, could be dependent on the fact that only 3 participants had such a level of impairment.

Interestingly, the model consistently achieved lower mRMSE values for HC compared to ALS participants across all modalities. This is confirmed by the values of the CV that for each model and modality is lower for HC. This suggests that predictions were more precise for HC, while for ALS higher variability in speech impairment seemed to make outcome prediction more challenging. Therefore, potential differences in data variability between HC and ALS participants could influence the model's generalization ability.

Based on the performance in the three domains, the video modality showed lower accuracy in predicting the outcome compared to the audio-based and multimodal approaches. This suggests that, for relatively low levels of speech impairment, video features are less informative in predicting the clinical outcome. This could be related to the lack of participants with severe speech impairment in the Toronto Neuroface dataset. Indeed, literature demonstrated that the analysis of kinematics could be a precursor of bulbar decline. Moreover, predictions based on the multimodal approach showed generally better performance compared with those from the models based exclusively on audio features. This may suggest that the integration of acoustic and kinematics features may provide a more comprehensive understanding of ALS conditions, by offering complementary information that could describe non-verbal aspects relevant to the ALS condition.

Although several studies have investigated using only audio [14], [15] or video [16] features as a tool for clinical outcome classification or prediction, only recently research started focusing on a multimodal approach [11], [19], achieving promising results. The features we used in both domains have already proved to be informative to this problem, but to the best of our knowledge, no studies exploited them for regression purposes. Thus, our work contributes to supporting and promoting further research on integrating audio and video analysis for ALS assessment. It also promotes the use of commonly adopted and low-cost instrumentations, which are highly available and feasible in both clinical settings and home environments, considering the possibility of using a smartphone for audio-video recording to enable remote monitoring, that can significantly improve the feasibility of large-scale, longitudinal studies, making them more accessible for patients and more scalable for researchers.

This study has some limitations. The small sample size, the limited number of ALS participants with relatively high speech impairment scores, and the restricted set of analyzed speech tasks constrained the results and their interpretation. Moreover, since some participants did not have both audio and video recordings, the three modalities are not perfectly comparable due to different subjects and instances. To address these limitations and expand the dataset, we analyzed each repetition as a separate instance. Despite these limitations, our results encourage future research on multimodal approaches for assessing orofacial functions in a non-invasive manner in neurological diseases.

Future work should explore additional regression models, larger datasets, and include more individuals with greater variability in speech impairment. Additionally, future studies should consider a broader range of relevant features and tasks, and longitudinal predictive studies, as well as enhance the methodology by incorporating additional steps for feature selection and statistical analyses.

## 5. Conclusions

In this work, we presented three different approaches to highlight the most suitable method for implementing a low-cost, non-invasive, and objective evaluation of speech impairment based on audio-video recordings. In particular, the best regression model was identified with the SVR, trained using multimodal features, achieving a mRMSE of 0.93 on a scale ranging from 5 to 25.

The approach that we proposed aims to support clinicians in assessing speech impairment, which is currently based on subjective and non-standardized methods, often limiting and delaying appropriate and rapid intervention to slow the decline of speech function and generally the quality of life. Indeed, an approach that integrates both audio and video analysis can capture several objective aspects from different domains, providing relevant and informative insights to enhance the understanding of the disease

## 6. Acknowledgements

The financial support of AriSLA – *Fondazione Italiana di ricerca per la SLA* is acknowledged (Project MIMOSA - Multimodal Intelligent Methods for Orofacial and Speech Assessment to predict ALS bulbar decline). The research in this paper uses the Toronto NeuroFace Dataset collected by Dr. Yana Yunusova and the Vocal Tract Visualization and Bulbar Function Lab teams at UHN-Toronto Rehabilitation Institute and Sunnybrook Research Institute respectively, financially supported by the Michael J. Fox Foundation, NIH-NIDCD, Natural Sciences and Engineering Research Council, Heart and Stroke Foundation Canadian Partnership for Stroke Recovery and AGE-WELL NCE.